\documentclass[a4paper,11pt]{article}
\usepackage{amsmath,amsthm,amssymb,graphicx}
\newcommand{\nc}{\newcommand}
\nc{\rnc}{\renewcommand}
\nc{\nn}{\nonumber}
\nc{\der}{{\partial}}
\rnc{\Im}{{\textrm{Im}\,}}
\rnc{\Re}{{\textrm{Re}\,}}
\nc{\db}{\displaybreak[0]\\}
\nc{\bra}{\langle}
\nc{\ket}{\rangle}
\nc{\astl}{\underset{\mathcal{L}_l}{\ast}}
\nc{\astj}{\underset{\mathcal{L}_j}{\ast}}
\nc{\astll}{\underset{\mathcal{L}_1}{\ast}}
\nc{\Yt}{Y^{\textrm{th}}}
\nc{\rmd}{{\rm d}}
\nc{\rmi}{{\rm i}}
\nc{\rme}{{\rm e}}

\DeclareMathOperator{\asec}{arcsec}

\nc{\End}{\mathrm{End}}

\begin{document}

\title{Numerical study on a crossing probability for the 
four-state Potts model: Logarithmic correction to the 
finite-size scaling}

\author{
Kimihiko Fukushima\thanks{2213091@alumni.tus.ac.jp} \, and
Kazumitsu Sakai$^2$\thanks{E-mail: k.sakai@rs.tus.ac.jp} \,
\\\\
\textit{Department of Physics,
Tokyo University of Science,}\\
 \textit{Kagurazaka 1-3, Shinjuku-ku, Tokyo, 162-8601, Japan} \\
\\\\
\\
}

\date{March 31, 2019}

\maketitle

\begin{abstract}
A crossing probability for the critical four-state Potts model on 
an $L\times M$ rectangle on a square lattice is numerically studied. 
The crossing probability here denotes the probability that spin clusters 
cross from one side of the boundary to the other.
First, by employing a Monte Carlo method, we calculate the fractal dimension of 
a spin cluster interface with a fluctuating boundary condition. By comparison
of the fractal dimension with that of the Schramm-Loewner evolution (SLE), we
numerically confirm that the interface can be described by the SLE with $\kappa=4$, 
as predicted in the scaling limit.
Then, we compute the crossing 
probability of this spin cluster interface for various system sizes and aspect ratios. 
Furthermore, comparing with the analytical results for the scaling 
limit, which have been previously obtained by a combination of the SLE 
and conformal field theory, we numerically find that the crossing probability 
exhibits  a logarithmic correction $\sim 1/\log(L M)$ to the finite-size scaling.
\end{abstract}

\maketitle
\noindent
1. {\it \large Introduction}\, The geometrical description of critical phenomena
has renewed interest in the theoretical study of phase transitions.
Various theoretical tools have been developed, particularly in the two-dimensional (2D) case 
where the structure of critical phenomena is expected to be 
strongly constrained by an infinite-dimensional conformal symmetry.

In particular, the Schramm-Loewner evolution (SLE)
\cite{S1,KN,C1,BB1,Gruz} describes the 
random fractals arising in 2D critical phenomena as a growth process 
defined by a stochastic evolution of conformal maps: the SLE generates
a random curve on a planar domain from a 1D Brownian motion on the
boundary. For instance, the SLE defined on the complex upper 
half-plane $\mathbb{H}$, which is conventionally called the chordal SLE,
is described by the evolution of the following conformal map:
\begin{equation}
dg_t(z)=\frac{2dt}{g_t(z)-\sqrt{\kappa}B_t}, \quad g_0(z)=z\in\mathbb{H},
\label{sle}
\end{equation}
where $B_t$ is the 1D standard Brownian motion living on $\mathbb{R}$,
i.e., its expectation value and variance are, respectively, 
given by ${\bf E}[dB_t]=0$ and ${\bf E}[dB_t dB_t]=dt$.
The tip $\gamma_t$ of the random curve evolves as 
$\gamma_t=\lim_{\epsilon\to+0}g^{-1}_t(\sqrt{\kappa}B_t+i\epsilon)$.
Namely, the SLE is able to give a direct description of \textit{non-local}
geometrical objects which are difficult to describe within traditional
approaches. The SLE has only one parameter $\kappa$ associated with
the diffusion constant of the Brownian motion. This parameter $\kappa$
qualitatively and quantitatively characterizes the random curves
generated by the SLE. For example, the fractal dimension $d_{\rm f}$ of 
the SLE curve is expressed as \cite{RS, Bef}:
\begin{equation}
d_{\rm f}=\min\left(2,1+\frac{\kappa}{8} \right).
\label{FD}
\end{equation}

On the other hand, the universal properties of 2D critical systems 
are explained by conformal field theory (CFT) \cite{BPZ,FMS}.
 In contrast to the SLE,
CFT determines correlation functions among \textit{local} operators. 
The SLE with $\kappa$ is conjecturally related to CFT \cite{BB2,FW1} via
\begin{equation}
c=\frac{(3\kappa-8)(6-\kappa)}{2\kappa},
\label{kappa}
\end{equation}
where $c$ is the central charge characterizing
the universality class of 2D critical systems.
By combining the SLE with CFT, we can systematically
analyze geometrical objects for  2D critical statistical 
mechanics models.

Crossing probabilities in statistical mechanics models, which are mainly 
discussed in this letter, are one of these non-local geometrical objects.
The crossing probability in this letter denotes  the probability
that a cluster composed by local variables such as spins
connects two disjoint segments on the boundary of a simply 
connected planar domain. Thanks to the conformal invariance expected in
the critical systems, this problem can be mapped to that of
the upper half-plane $\mathbb{H}$. 
For the statistical models with cluster boundaries described
by the SLE with $\kappa>4$, the crossing probability 
can be computed by use of the single SLE curve \eqref{sle}.
The result contains the one for critical
percolation ($\kappa=6$; $c=0$), which is well-known as the Cardy formula
\cite{Card2}, and for the Fortuin-Kasteleyn (FK) clusters
in the Ising model ($\kappa=16/3$; $c=1/2$). On the other hand,
the crossing probability for the interfaces characterized by the
SLE with $\kappa\le 4$, which includes the spin cluster boundaries of the
Ising model ($\kappa=3$; $c=1/2$), may be evaluated by considering the multiple
(three) SLE curves as constructed in \cite{BBK}. 

There still exists, however, a non-trivial problem:
what kinds of cluster boundaries in the scaling limit of the lattice model 
do SLE curves actually correspond to? This problem becomes serious for 
the systems possessing multiple (typically more than two) local variables, 
such as the three- or four-state Potts models (see, for instance, \cite{GC} 
for a treatment of the three-state Potts model), where various kinds of cluster
boundaries can be defined.

In this letter, using a Monte Carlo method, we numerically compute 
the crossing probability for certain spin clusters of the  four-state
Potts model on an $L\times M$ rectangle on a square lattice.
In the scaling limit, the universal behavior of the four-state Potts model 
is classified by the $c=1$ CFT. Intuitively, from \eqref{kappa}, appropriately 
defined cluster boundaries become the SLE curves with $\kappa=4$ in the scaling limit.
Indeed, it is known that the FK clusters of the model become the SLE 
with $\kappa=4$ (see \cite{KN,BB1,Gruz} for example).
We hypothesize that properly defined spin cluster boundaries
are also described by the SLE with $\kappa=4$. First, we numerically 
evaluate the fractal dimension of a spin cluster interface with a 
`fluctuating' boundary condition. Comparing the result with that 
obtained by the SLE,  we numerically confirm that the above hypothesis
is correct for the interface with the fluctuating boundary condition. 
Second, from high-precision Monte Carlo data, 
we find that the spin cluster interface under the fluctuating
boundary condition gives the crossing probability described by 
the multiple version of the SLE.
Moreover, comparing the results with those obtained in the 
scaling limit \cite{BBK}, we numerically show that the crossing probability exhibits  
a logarithmic correction $\sim 1/\log(L M)$ to the 
finite-size scaling.\\

\noindent
2. {\it \large Potts Model}\,
We shall study the  Potts model
on a rectangle on a square lattice. The $q$-state Potts model
is an interacting spin model with the spin variables taking on the
values $1,2,\dots,q$. The partition function $Z$ of the system
is given by
\begin{equation}
Z=\sum_{\{\sigma_j\in Q\}}
\exp\left[J \sum_{\bra j,k \ket}\delta_{\sigma_j,\sigma_k} \right],
\quad Q=\{1,2,\dots, q\},
\label{Potts}
\end{equation}
where $J>0$ (i.e., ferromagnetic interaction) and $\bra j, k \ket$
denotes adjacent sites on the square lattice. For $q\to 1$ and
$q=2$, the model is equivalent to the bond percolation model
and the Ising model, respectively. It is well known that 
the Potts model exhibits
a second-order phase transition for $q\le 4$, while for $q>4$ the
transition is of first order. Note that, for $q\le 4$, the scaling 
behavior is governed by  CFT \cite{VF} with
\begin{equation}
c=1-\frac{6}{s(s+1)}, \quad
s=-1+\frac{\pi}{\asec(2/\sqrt{q})}.
\label{q-c}
\end{equation}
\begin{table}[t]
\centering
  \begin{tabular}{|l||l|l|} \hline
    $q$ & $c$ & $\kappa$ \\ \hline  \hline
    $1$ &$0$  & $8/3,6$ \\ \hline
    $2$ &$1/2$&$3,16/3$ \\ \hline
    $3$ &$4/5$&$10/3,24/5$\\ \hline
    $4$ &$1$  &$4$ \\ \hline
  \end{tabular}
\caption{The values of $\kappa$ and $c$  calculated by \eqref{kappa} and \eqref{q-c} 
for the $q$-state Potts model. The relations for $q=1$ (critical percolation) \cite{Smi01,Smi10} 
and  for $q=2$ (Ising) \cite{CDHKS} have been rigorously proven.}
\label{kappa-c}
\end{table}
In Table \ref{kappa-c}, the values of $\kappa$ and $c$ calculated by \eqref{kappa} and \eqref{q-c} 
for the $q$-state Potts model 
are summarized (see \cite{KN,BB1,Gruz} and references therein 
for some examples of the correspondence between SLE and interfaces
in the scaling limit of some lattice models including the FK clusters 
of the Potts model). The relations $\kappa=8/3, 6$ for $q=1$ (critical percolation) 
\cite{Smi01,Smi10} 
and $\kappa=3,16/3$ for $q=2$ (Ising) \cite{CDHKS} have been rigorously proven.
The border value $q=4$ (corresponding to $c=1$ and $\kappa=4$) 
is the most crucial: though the model undergoes 
a second-order transition,
the power-law behavior of local operators is expected to be  
modified by logarithmic factors from marginally irrelevant 
operators \cite{NS,CNS,SS,AA,CZ}. For our purpose, hereafter
we set $J=J_{\rm c}:=\log(\sqrt{q}+1)$ where $J_{\rm c}$
denotes the critical temperature.
%, and mainly focus on the
%case $q=4$, i.e., $\kappa=4$ unless otherwise specified.
\\

\noindent
3. {\it \large Multiple SLE and
Crossing Probabilities}\, The SLEs can be generalized to 
multiple versions generating $N$ random curves
evolving from $N$ points on the boundary as in 
\cite{BBK, Card, Dub, Gra, Koz,Sak}.
For the chordal case, the multiple SLE is given by
the following form \cite{BBK}:

\begin{equation}
dg_t(z)=\sum_{j=1}^N \frac{2 dt}{g_t(z)-X_t^{(j)}},
\quad
d X_t^{(j)}=\sqrt{\kappa} dB_t^{(j)}+d F_t^{(j)},
\end{equation}
where $\{B_t^{(j)}\}$ is an $\mathbb{R}^{N}$-valued Brownian motion
whose expectation value and variance are given by
${\bf E}[dB^{(j)}_t]=0$ and ${\bf E}[dB^{(j)}_t dB^{(k)}_t]=\delta_{jk}dt$, respectively.
Reflecting interactions among curves, the driving term $X_t^{(j)}$ has an additional
drift term $dF_t^{(j)}$ given by
\begin{equation}
dF_t^{(j)}=\kappa dt (\der_{X_t^{(j)}}\log Z_t)+\sum_{k\ne j} \frac{2dt}{X_t^{(j)}-X_t^{(k)}}.
\end{equation}
Here $Z_t$ is a correlation function  of boundary condition 
changing (bcc) operators:
\begin{equation}
Z_t=\left\bra \psi(\infty)\psi_{2,1}(X_t^{(1)})\cdots \psi_{2,1}(X_t^{(N)}) \right\ket,
\end{equation}
where the $\psi_{2,1}(X_t^{(j)})$ are bcc operators inserted at the positions 
$z=X_t^{(j)}$ and
are degenerate at level two: the conformal weight $h$ of $\psi_{2,1}$ is given by
$h=h_{2,1}=(6-\kappa)/(2\kappa)$.  Also $\psi(\infty)$ is a bcc operator
inserted at $z=\infty$. The tip $\gamma_t^{(j)}$ 
of the $j$th curve is given by $\gamma_t^{(j)}=\lim_{\epsilon\to+0}g_t^{-1}(X_t^{(j)}+i\epsilon)$.
Most importantly, the conformal weight of  
$\psi(\infty)$ (denoted by $h_{\infty}$) characterizes 
the topological configurations of the SLE curves \cite{BBK}.
This indicates that some \textit{non-local} geometrical 
properties can be determined by the expectation value of 
products of \textit{local} operators, which can be exactly computed by
CFT. 

\noindent
\begin{figure}[t]
\centering
\includegraphics[width=0.6\textwidth]{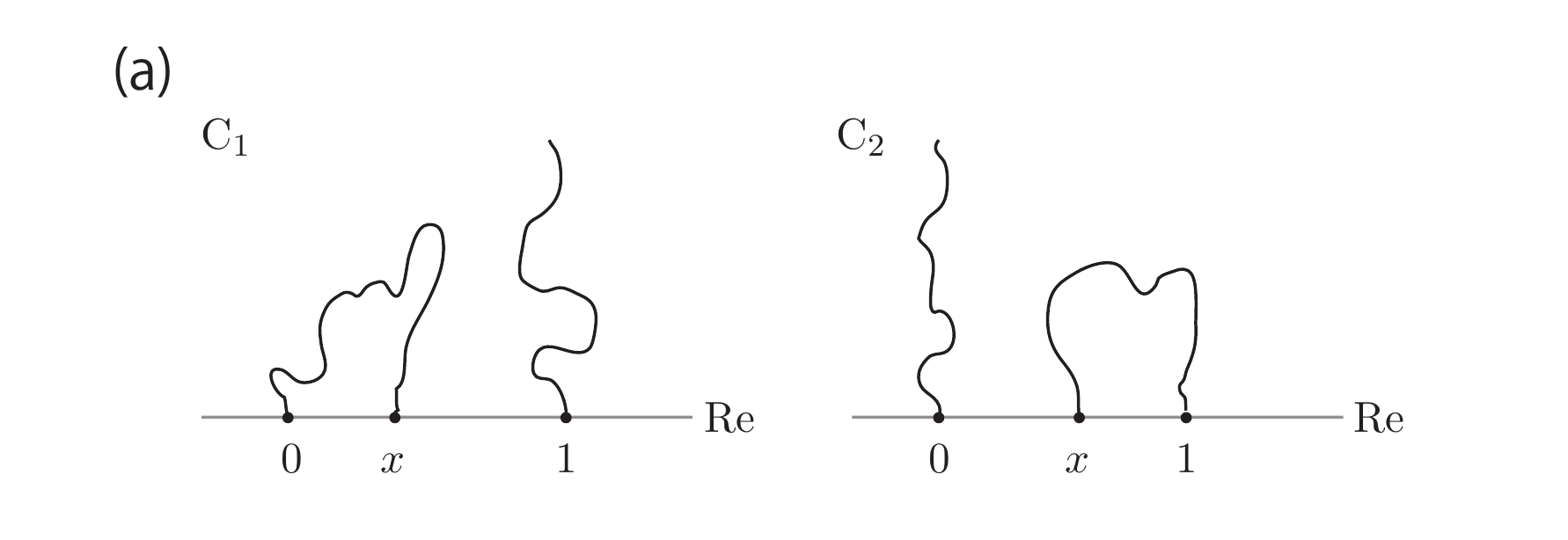}
\includegraphics[width=0.6\textwidth]{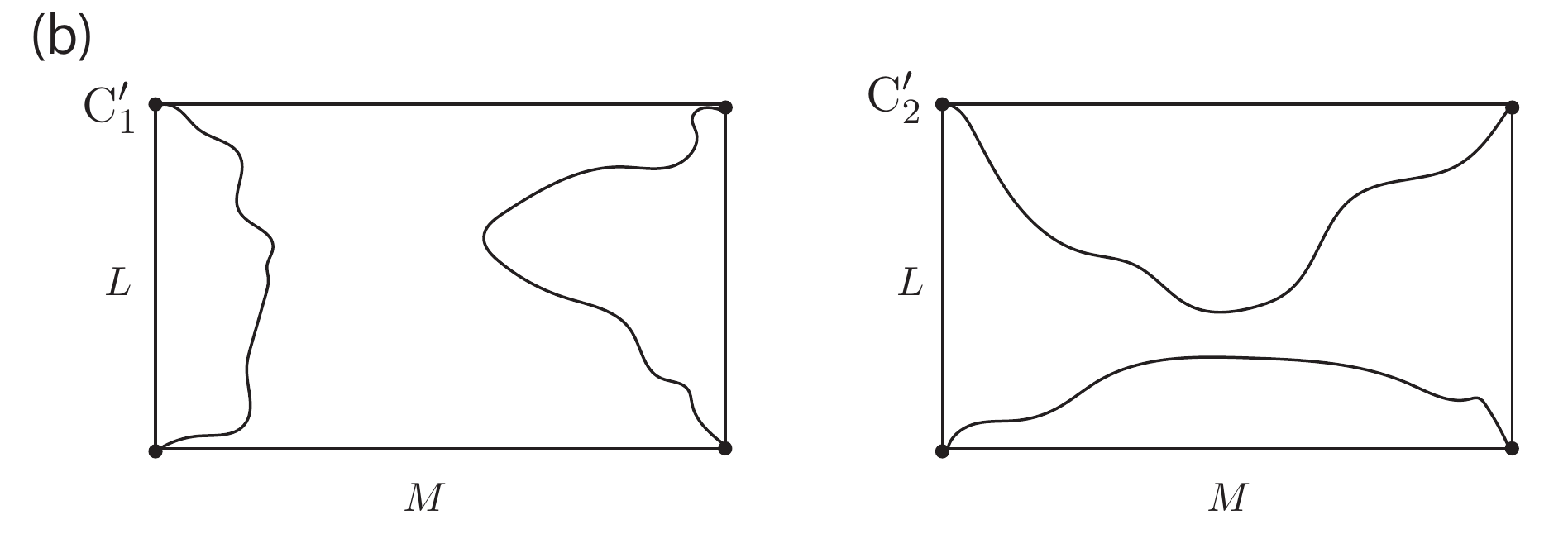}
\caption{(a) The two topologically inequivalent configurations of the three 
SLE curves
starting from the points $z=0$, $x$ ($0<x<1$), $1$. ${\rm C_1}$ (resp. ${\rm C_2}$) 
denotes the two curves evolving from $z=0$ (resp. $z=1$) and $z=x$ hit 
each other, and the curve starting from $z=1$ (resp. $z=0$) eventually 
converges to infinity. (b) The two configurations ${\rm C_1}$ and 
${\rm C_2}$ in (a)
are mapped by the Schwarz-Christoffel transformation to those 
defined on a rectangle with the aspect ratio $r=L/M$ as in 
${\rm C'_1}$ and ${\rm C'_2}$, respectively. The relation between $x$ in 
(a) and $r$ is determined by \eqref{asp}. The configuration ${\rm C'_1}$ 
corresponds to Fig. \ref{Ising} (a) or Fig. \ref{Potts-config} (c) 
that the spin clusters, which consist 
of the spin variables same as those on the top edge, 
cross from the top edge to the bottom edge.} 
\label{config}
\end{figure}
\begin{figure}[htb]
\centering
\includegraphics[width=0.6\textwidth]{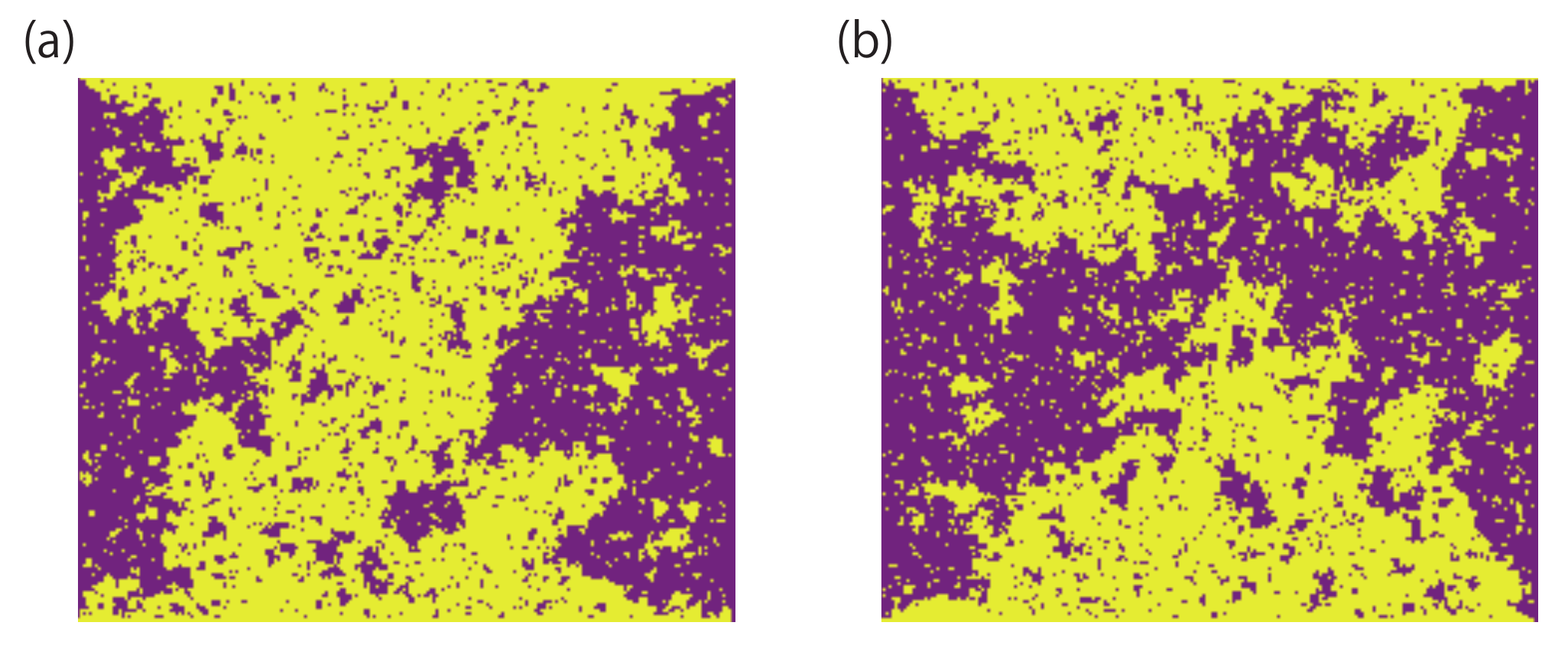}
\includegraphics[width=0.6\textwidth]{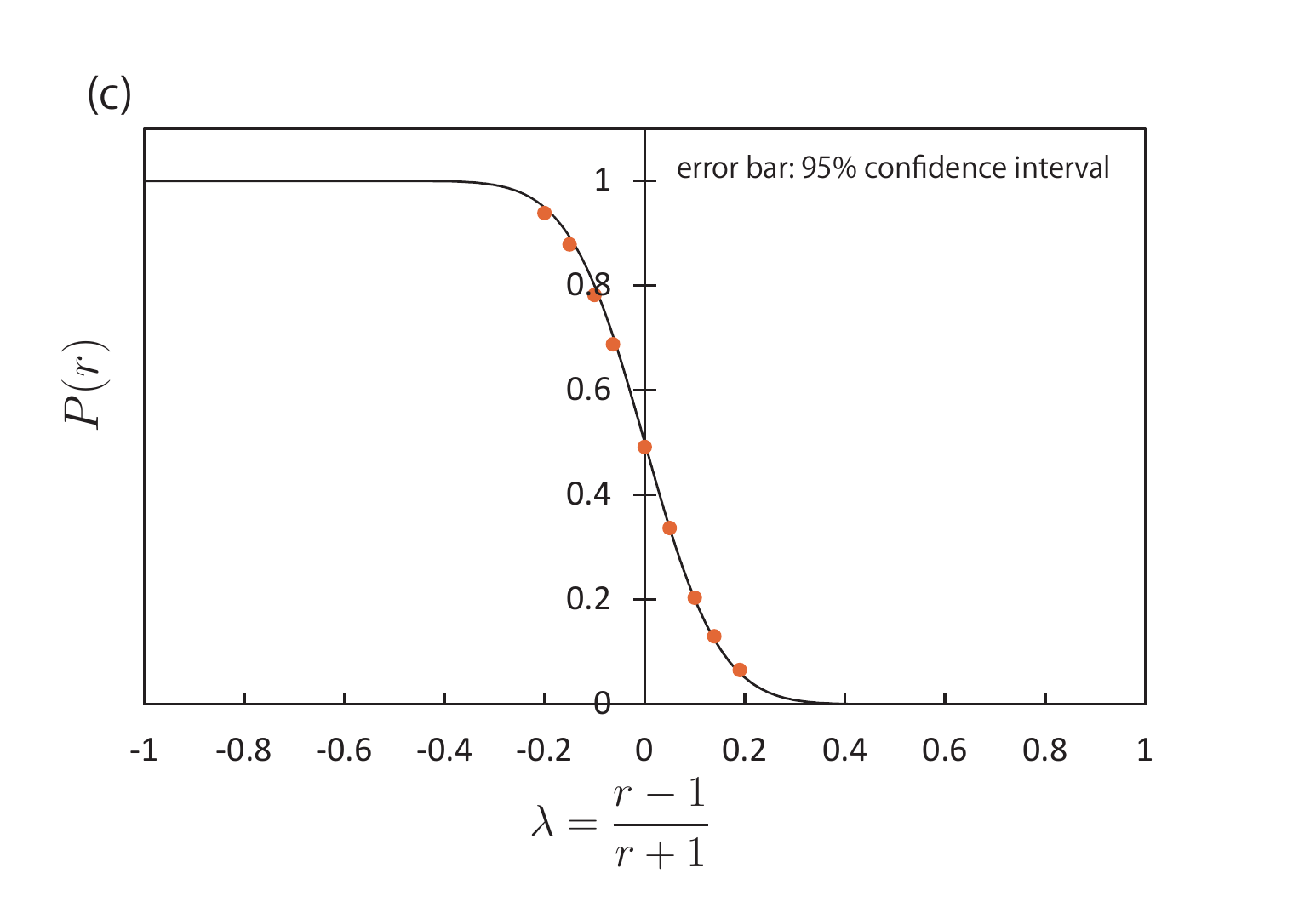}
\vspace{-11pt}
\caption{Upper panel: Typical snapshots of the Ising model
where the boundary spins on the top and bottom edges (resp. left and right edges)
are fixed to  1 (colored yellow) (resp. 2 (colored purple)).
The spin cluster boundaries (i.e. the boundaries of yellow and purple clusters)
starting from the corners  are described by the multiple SLE curves as schematically shown in Fig.\ref{config} (b).  
The configurations ${\rm C_1'}$ and ${\rm C_2'}$
correspond to (a) and (b), respectively. 
Lower panel: Numerical calculation 
of the crossing probability of the occurrence of configuration (a) for
the case $L\times M=1,600$. The data coincides with the one
 in the scaling limit depicted as the thick line, 
which is analytically given by \eqref{prob2} with 
$\kappa=3$. The error bars are smaller than the point size used.}  
\label{Ising}
\end{figure}
\begin{figure}[htb]
\centering
\includegraphics[width=0.6\textwidth]{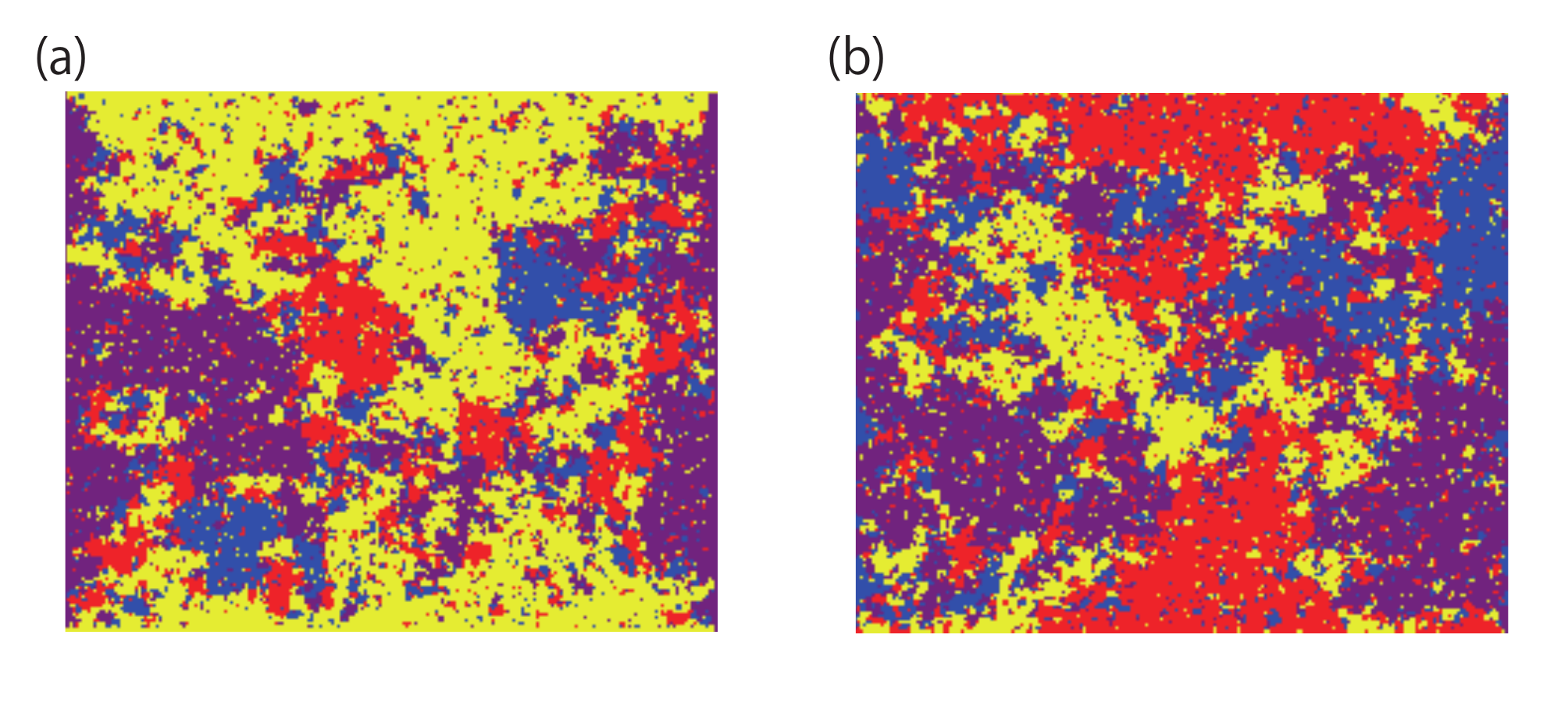}\\[4mm]
\includegraphics[width=0.6\textwidth]{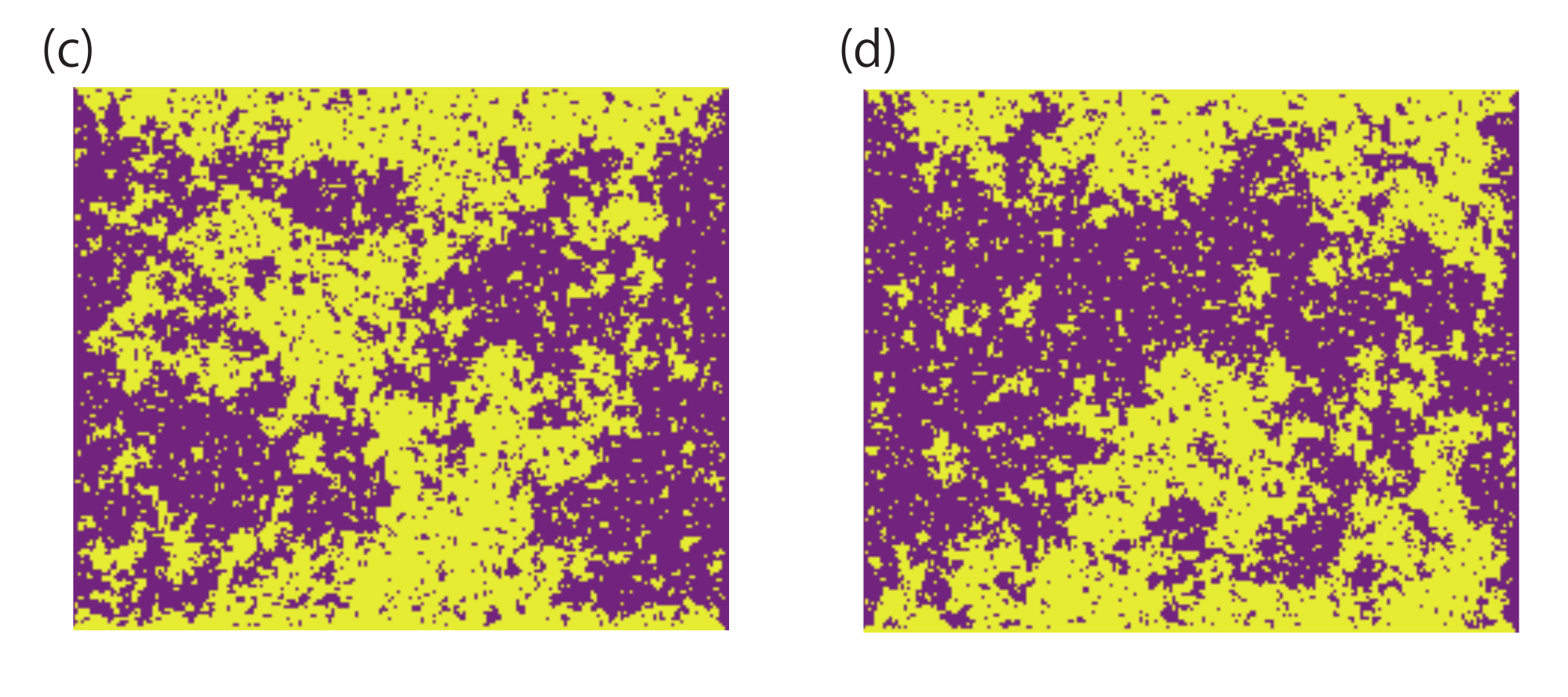}
\caption{Upper panel: (a) A typical snapshot of the four-state Potts model
where the boundary spins on the top and bottom edges (resp. left and right edges)
are fixed to  1 (colored yellow) (resp. 3 (colored purple)). The other
spins, i.e., type 2 and 4 are colored red and blue, respectively. The spin cluster 
boundary (i.e. the boundary between  yellow and purple clusters) starting 
from one corner  splits into two different types
of spin cluster interfaces (e.g. red-purple and red-yellow interfaces), 
when it encounters the other spin clusters (e.g. red clusters) in the bulk. 
(b) A snapshot of the four-state Potts model
with the fluctuating boundary condition. On the top and bottom edges 
(resp. left and right edges), the spins of type 1 and 2
(resp. 3 and 4) are randomly assigned. Lower panel: Snapshots
of the four-state Potts model under the fluctuating
boundary condition where the spins of  $\{1,2\}$ are 
colored yellow, while the spins of $\{3,4\}$ are
colored purple.  The configurations (c) and (d),
respectively, correspond to ${\rm C_1'}$ and ${\rm C_2'}$
in Fig.\ref{config} (b). 
}  
\label{Potts-config}
\end{figure}

For instance, for $N=3$ and $h_{\infty}=h_{2,1}$,  the two topologically 
inequivalent configurations are allowed. See Fig.~\ref{config} (a) for these
configurations where the three curves are assumed to start from the points
$X_{t=0}^{(1)}=0$, $X_{t=0}^{(2)}=x$ ($0<x< 1$) and $X_{t=0}^{(3)}=1$. Each
configuration is characterized by the correlation function
\begin{equation}
Z_0=\bra \psi_{2,1}(\infty)\psi_{2,1}(0)\psi_{2,1}(x)\psi_{2,1}(1) \ket
=Z_{\rm C_1}(x)+Z_{\rm C_2}(x).
\label{cor}
\end{equation}
 Namely, the probability ${\bf P}[{\rm C}_1]$
(resp. ${\bf P}[{\rm C}_2]$)
of the occurrence of the configuration $\rm {C}_1$ (resp. $\rm{C}_2$) 
as in the left (resp. right) panel of Fig. \ref{config} (a) is expressed as
\begin{equation}
{\bf P}[{\rm C}_1]=\frac{Z_{\rm C_1}(x)}{Z_{\rm C_1}(x)+Z_{\rm C_2}(x)},
\quad
{\bf P}[{\rm C}_2]=\frac{Z_{\rm C_2}(x)}{Z_{\rm C_1}(x)+Z_{\rm C_2}(x)},
\label{prob}
\end{equation}
where up to an overall factor $Z_{\rm C_1}(x)$ and  $Z_{\rm C_2}(x)$ 
are, respectively, given by \cite{BBK}:
\begin{align}
&Z_{\rm C_1}(x)=(1-x)^{2/\kappa} x^{2/\kappa}
{}_2 F_1\left(\frac{4}{\kappa},\frac{12-\kappa}{\kappa};\frac{8}{\kappa};1-x\right),\nn \\
&Z_{\rm C_2}(x)=Z_{\rm C_1}(1-x).
\label{Z}
\end{align}
By the Schwarz-Christoffel transformation,
the two configurations ${\rm C_1}$ and ${\rm C_2}$ defined on 
$\mathbb{H}$ can be mapped to those on a rectangle (see Fig. \ref{config} (b)) 
with the aspect ratio $r$:
\begin{equation}
r=\frac{K(1-k^2)}{2K(k^2)},\quad k=\frac{1-\sqrt{x}}{1+\sqrt{x}},
\label{asp}
\end{equation} 
where $K(k^2)$ is the complete elliptic integral of the first kind with 
the modulus $k$ (see, for instance, Chapter 11 in \cite{FMS} 
for a detailed derivation, and see also arguments in
\cite{FK,FSKZ} 
for more general polygons). The probability of the occurrence
of the configuration ${\rm C'_1}$ in the left panel of Fig. \ref{config} (b)
corresponds to the  crossing probability that the spin clusters cross from 
the top edge to the bottom edge (see also Fig. \ref{Ising} (a) or Fig. \ref{Potts-config}
(c) for a configuration of spin clusters corresponding 
to ${\rm C'_1}$). In this letter, we discuss
this crossing probability defined  on the rectangular domain. 
By an assumption that the probabilistic
measure is invariant under the conformal transformation,
the crossing probability for the rectangle of aspect ratio $r$
is given by 
\begin{equation}
P(r)={\bf P}[\rm C_1]
\label{prob2}
\end{equation} 
with \eqref{prob}, \eqref{Z} 
and \eqref{asp}.

For the four-state Potts model ($\kappa=4$), the probability reduces to 
a simpler form ${\bf P}[{\rm C_1}]=1-x$ and ${\bf P}[{\rm C_2}]=x$,
which can be directly obtained by setting $\kappa=4$ in \eqref{Z}.
Thus, the analytical expression
of the crossing probability in the scaling limit
of the four-state Potts model explicitly reads
\begin{equation}
P(r)=1-x \quad (0<x<1),
\label{scaling}
\end{equation}
where $x$ is a function of $r$ as given by \eqref{asp}.
%$P(r)$ against the parameter 
%$\lambda=(r-1)/(r+1)$ is plotted in Fig.\ref{crossing}.
\\

\noindent
4. {\it \large Numerical Results}\, Now, let us numerically calculate the crossing
probability of the four-state Potts model \eqref{Potts} on a rectangle
on the square lattice.
 First we shall detect the interface and the boundary condition 
giving the above $P(r)$ \eqref{scaling}. In fact, several types 
of spin cluster boundaries can be defined in the Potts model for $q\ge 3$, 
which is in contrast to the Ising case ($q=2$) where there
exists only one type of spin cluster boundary: the interfaces
between clusters of spin type 1 and those of spin type 2.
For the Ising model, we therefore can easily set up a
boundary condition of the rectangle which is compatible with 
the multiple SLE with $\kappa=3$. Namely, the boundary spins 
on the top and bottom edges are set to type 1, and
those on the left and right edges are fixed to type 2 and vice versa
(see Fig.~\ref{Ising} for an example). 
Indeed, as in Fig.~\ref{Ising}, the numerical results calculated by a 
Monte Carlo simulation (the Wolf algorithm) for the crossing  
probability agree well with the ones in the scaling limit, which is
obtained by \eqref{prob2} with $\kappa=3$.

Let us go back to the case of the four-state Potts model ($q=4$; $\kappa=4$).
The situation is quite different from the Ising  case.
For instance, if we fix the boundary spins as for the Ising model
(see Fig. \ref{Potts-config} (a)), the spin cluster boundaries starting 
from the corners no longer join any corners, due to the 
existence of clusters consisting of spins 
different from the ones on the boundaries. Namely, the spin cluster 
boundary starting from one corner splits into two different types
of spin cluster interfaces, when it meets the other spin clusters
in the bulk, which is not compatible with the SLE  (see a snapshot in 
Fig. \ref{Potts-config} (a)).

Alternatively, we adopt  a `fluctuating' boundary condition
as defined in \cite{GC,FSKZ}. The fluctuating boundary condition
in our case is defined as follows. We divide the four types of spin 
$1,2,3,4$ into two parts, say $\{1,2\}$ and $\{3,4\}$ (see 
\cite{FSKZ} for another choice of the fluctuating boundaries). On the top and bottom edges (resp. left and right edges), 
the  type 1 and 2 spins (resp. the type 3 and 4 spins) are randomly assigned
and vice versa. See Fig. \ref{Potts-config} (b) as an example. Under this configuration,
the spin cluster boundary is defined as the interface between the spin 
cluster consisting of $\{1,2\}$ and the one consisting of $\{3,4\}$.
See  Fig. \ref{Potts-config} (c) and (d) where the type $\{1,2\}$
is colored yellow, while the type $\{3,4\}$ is colored purple.
Consequently, we can construct the spin cluster boundary
compatible with the SLE.

\begin{figure}[htb]
\centering
\includegraphics[width=0.6\textwidth]{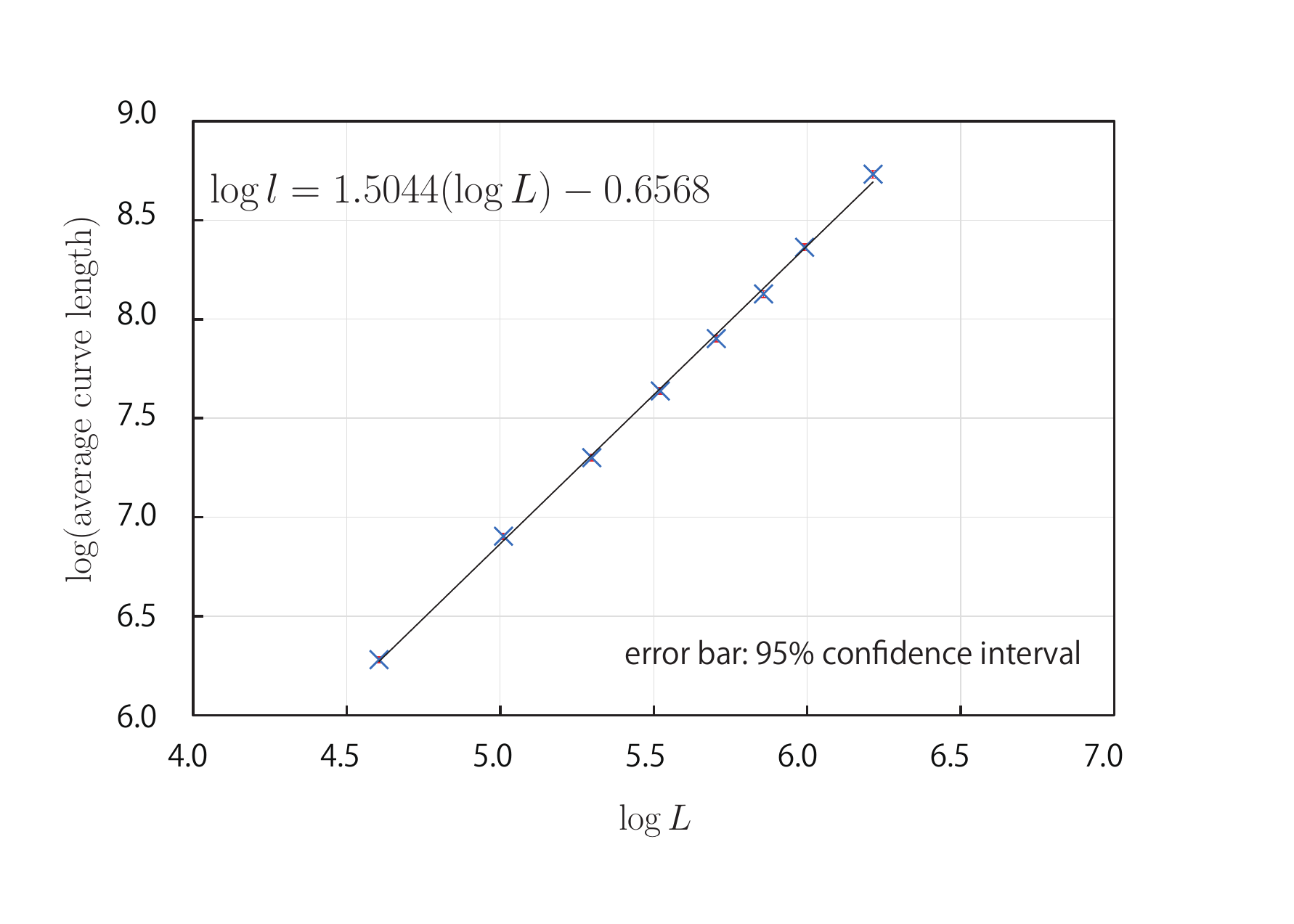}
\caption{The numerical evaluation of the fractal dimension of the
spin cluster boundary of the four-state Potts model on
the rectangle with the fluctuating boundary condition.
The fractal dimension $d_{\rm f}$ is evaluated as $d_{\rm f}=1.504\pm 0.020$
by the least squares method. This value is consistent with the prediction
from the SLE \eqref{FD} with $\kappa=4$: $d_{\rm f}=3/2$. The error bars
are smaller than the point size used.}  
\label{fractal}
\end{figure}

Indeed, by numerically analyzing the
fractal dimension,   we can confirm that this spin cluster 
boundary can be described by the SLE with $\kappa=4$. Here the fractal 
dimension $d_{\rm f}$ is evaluated in the following scaling law:
\begin{equation}
l\sim a\left(\frac{L}{a}\right)^{d_{\rm f}} \quad (L/a\gg 1),
\end{equation}
where $l$ is the total length of the curve, and $L$ and $a$ are the system 
size and the lattice spacing, respectively.
As shown in Fig. \ref{fractal}, using a Monte Carlo method (the Wolf algorithm)
we numerically confirm that the fractal dimension
of the above-defined spin cluster boundary is $d_{\rm f}=1.504\pm 0.020$,
which is obtained by the least squares method. 
The result agrees well with the prediction from the SLE of $d_{\rm f}=3/2$
obtained by the substitution of  $\kappa=4$ into the formula \eqref{FD}.
Note that, in the scaling limit, 
the boundary spins $1$ and $2$ (resp. $3$ and $4$) are uniformly distributed 
on the top and bottom edges (the left and right edges) for the fluctuating 
boundary condition. Therefore,  for our 
numerical calculations, we arrange the boundary spins alternately
as $1212\cdots$ (resp. $3434\cdots$) on the top and bottom edges
(resp. the left and right edges).
\begin{figure}[ttt]
\centering
\includegraphics[width=0.6\textwidth]{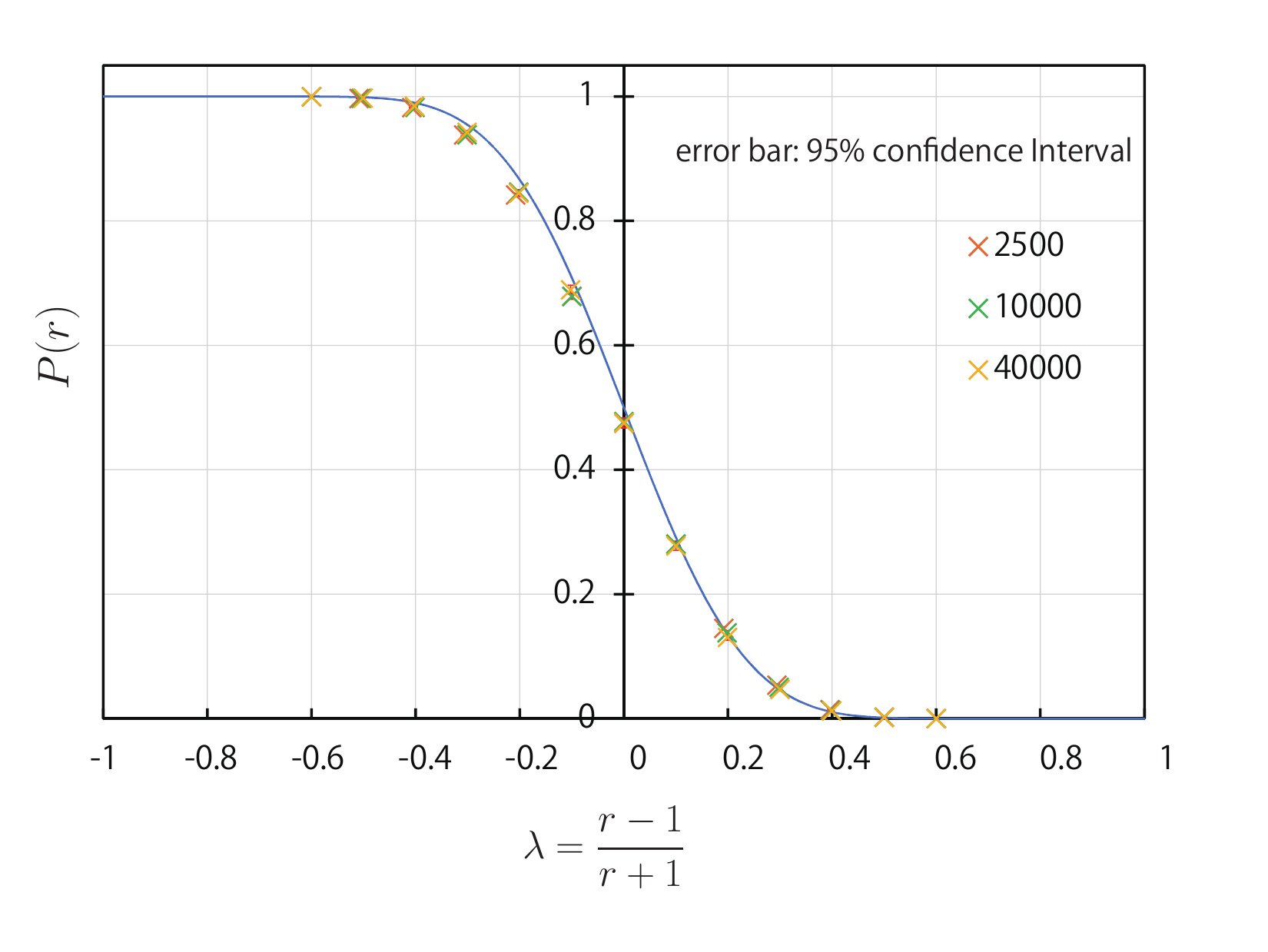}
\caption{Numerical computation
of the crossing probability of the spin cluster boundary
for the four-state Potts model on the rectangle with
the fluctuating boundary condition. Compared with the
Ising model as shown in Fig. \ref{Ising} (c), the numerical
data slowly converges to the analytical result \eqref{scaling} (and 
\eqref{asp})
which is depicted as the thick line.
}  
\label{cp-Potts}
\end{figure}

\begin{figure}[ttt]
\centering
\includegraphics[width=0.99\textwidth]{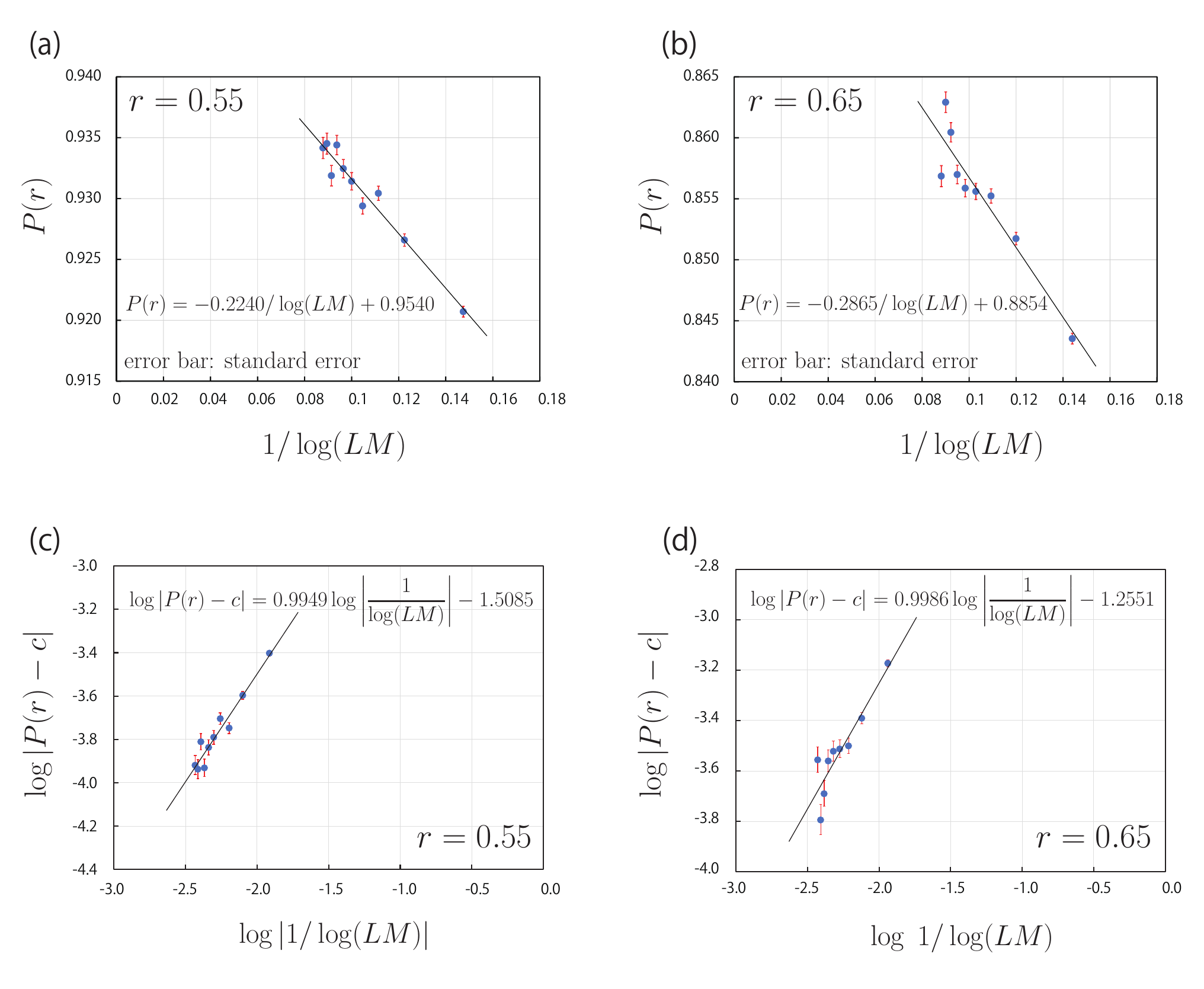}
\caption{The crossing probability 
$P(r)$ depicted against $1/\log (L M)$
for the fixed aspect ratio of $r=L/M=0.55$ (a) and $r=0.65$ (b),
which are  fitted to a line calculated by the linear least squares method.
The logarithmic behavior can be more clearly confirmed by
panels (c) and (d).}  
\label{log-cor}
\end{figure}

Next, we examine the crossing probability for the
four-state Potts model on the rectangle with the fluctuating
boundary condition.
Figure \ref{cp-Potts} shows the numerical results of the
crossing probability for various
system sizes and aspect ratios. In comparison with the 
Ising model in Fig. \ref{Ising} (c), the finite-size data
converges much more slowly to that in the scaling limit
given by \eqref{scaling}. This slow convergence might be
explained in terms of logarithmic corrections caused by
the existence of marginally irrelevant operators, since
 the crossing probability 
is essentially governed by the correlation function
of local operators
\eqref{cor}. In fact, as shown in Fig. \ref{log-cor} (a) 
and (b), the crossing probability for the $L\times M$ rectangle 
exhibits a logarithmic correction 
$\sim 1/\log(L M)$ to the finite-size scaling. 
This logarithmic behavior can be more clearly confirmed by 
 Fig. \ref{log-cor} (c) and (d), as both $\log|P(r)-c|$ 
are linearly dependent with $\log(\log LM)$. This behavior 
cannot be explained by power functions. Similar logarithmic behavior 
is also observed for various values of $r$.
This behavior also might be analytically confirmed by 
analyzing the correlation function \eqref{cor}.\\

\noindent
5. {\it \large Concluding Remarks}\, In this letter, we have numerically 
investigated the crossing probability of the Potts model on the 
$L\times M$ rectangle for the fluctuating boundary condition. 
Comparing  the fractal dimension
with the prediction from  the SLE,  we have numerically confirmed 
that the spin cluster interfaces under the fluctuating boundary condition
are described by the SLE with $\kappa=4$.
We also have shown numerically that the 
crossing probability of this spin cluster boundary exhibits a 
logarithmic correction  $\sim 1/\log(L M)$ to the finite-size scaling.
This logarithmic correction might be explained
by the existence of marginally irrelevant 
operators. Recently, similar logarithmic behavior has also 
been observed in other random geometries (four-point boundary
connectivities of FK clusters) for the four-state Potts model
\cite{GV18} (see also \cite{GV17} for the Ising model).
It remains a crucial problem  to show this logarithmic 
behavior analytically by analyzing the correlation function
of the local operators.
 
%
%%%%%%%%%%%%%%%%%%%%%%%%%%%%%%%%%%%%%%%%%%%%%%%%%%
\section*{Acknowledgment}
%%%%%%%%%%%%%%%%%%%%%%%%%%%%%%%%%%%%%%%%%%%%%%%%%%%%%%%%%%%%%%%%

The present work was partially supported by Grant-in-Aid for Scientific
Research (C) No. 16K05468 from Japan Society for the Promotion of Science.

\end{document}